%Paper: cond-mat/9401061
%From: Hao Li <haoli@chaos.uchicago.edu>
%Date: Wed, 26 Jan 94 14:46:52 CST

\magnification=\magstep1
\font\titlefont=cmr10 scaled\magstep3
\font\secfont=cmbx10 scaled\magstep1

%\font\titlefont=cmr10
%\font\secfont=cmr10
%-------------------------------------------------------------------------
\newbox\leftpage
\newdimen\fullhsize
\newdimen\hstitle
\newdimen\hsbody
\tolerance=1000\hfuzz=2pt
\hoffset=0.0truein \voffset=0.20truein
\hsbody=\hsize \hstitle=\hsize
\baselineskip=20pt plus 4pt minus 2pt
%---------------------------------------------------------------------
%use \nolabels to get rid of eqn and ref labels in draft mode
\def\nolabels{\def\eqnlabel##1{}\def\eqlabel##1{}\def\reflabel##1{}}
\def\writelabels{\def\eqnlabel##1{%
{\escapechar=` \hfill\rlap{\hskip.09in\string##1}}}%
\def\eqlabel##1{{\escapechar=` \rlap{\hskip.09in\string##1}}}%
\def\reflabel##1{\noexpand\llap{\string\string\string##1\hskip.31in}}}
\nolabels
%---------------------------------------------------------------------
\def\title#1{ \nopagenumbers\hsize=\hsbody%
\centerline{ {\titlefont #1} }%
\pageno=0}
\def\author#1{\vskip 1 truecm%
\centerline{{\sl #1}}%
\centerline{The James Franck Institute and Department of Physics}%
\centerline{The University of Chicago}%
\centerline{Chicago, IL 60637}
\vskip 1cm}
\def\abstract#1{\centerline{\bf ABSTRACT}\nobreak\medskip\nobreak\par #1}
\def\pacs#1{\medskip\noindent #1
\vfill\eject\footline={\hss\tenrm\folio\hss}}
%---------------------------------------------------------------------
% tagged sec numbers
\global\newcount\secno \global\secno=0
\global\newcount\meqno \global\meqno=1
\def\newsec#1{\global\advance\secno by1
\xdef\secsym{\ifcase\secno
\or I\or II\or III\or IV\or V\or VI\or VII\or VIII\or IX\or X\fi }
%\xdef\secsym{\the\secno}
\global\meqno=1
\bigbreak\bigskip%\fi% (combination \goodbreak\bigskip\bigskip)
\noindent{\secfont\secsym. #1}\par\nobreak\medskip\nobreak}
\xdef\secsym{}

\def\appendix#1#2{\global\meqno=1\xdef\secsym{\hbox{#1}}\bigbreak\bigskip
\noindent{\bf Appendix #1. #2}\par\nobreak\medskip\nobreak}
%---------------------------------------------------------------------
%       \eqn\label{a+b=c}       gives a displayed equation with number
%                               chosen consecutively within sections.
%     \eqnn and \eqna define labels in advance

\def\eqnn#1{\xdef #1{(\the\secno.\the\meqno)}%
\global\advance\meqno by1\eqnlabel#1}
\def\eqna#1{\xdef #1##1{\hbox{$(\the\secno.\the\meqno##1)$}}%
\global\advance\meqno by1\eqnlabel{#1$\{\}$}}
\def\eqn#1#2{\xdef #1{(\the\meqno)}\global\advance\meqno by1%
$$#2\eqno#1\eqlabel#1$$}
\def\meqn#1{\xdef #1{\the\secno.\the\meqno}\global\advance\meqno by1}
%---------------------------------------------------------------------
%                        footnotes
\global\newcount\ftno \global\ftno=1
\def\foot#1{{\baselineskip=14pt plus 1pt\footnote{$^{a}$}{#1}}%
\global\advance\ftno by1}
%---------------------------------------------------------------------
%     \ref\label{text}
% generates a number, assigns it to \label, generates an entry.
% To list the refs on a separate page,  \listrefs
\global\newcount\refno \global\refno=1
\newwrite\rfile
\def\ref#1#2{[\the\refno]\nref#1{#2}}
\def\nref#1#2{\xdef#1{[\the\refno]}%
\ifnum\refno=1\immediate\openout\rfile=refs.tmp\fi%
\immediate\write\rfile{\noexpand\item{#1\ }\reflabel{#1}#2}%
\global\advance\refno by1}
\def\addref#1{\immediate\write\rfile{\noexpand\item{}#1}}

%---------------------------------------------------------------------
%               and finally, figures:
\def\caption{\centerline{{\bf Figure Captions}}\medskip\parindent=40pt}
\def\fig#1#2{\medskip\item{Fig.~#1:  }#2}
%               and finally finally, tables:

%---------------------------------------------------------------------
\def\frac#1#2{{#1\over#2}}

\def\epc{\epsilon_c}
\def\tho{\theta_0}
\def\ths{\theta_<}
\def\thl{\theta_>}

\def\degs{^\circ}
%-------------------------------------------------------------------
\centerline{(submitted to Europhysics Letters)}
\medskip
\title{Singular Shape of a Fluid Drop}
\title{in an Electric or Magnetic Field}
\author{Hao Li, Thomas C. Halsey, and Alexander Lobkovsky}
\abstract{Beyond a threshold, electric or magnetic fields cause a
dielectric or ferromagnetic fluid drop respectively to develop
conical tips. We analyze the appearance of the conical tips and the
associated shape
   transition of the drop
 using a local force balance as well as a global energy
   argument. We find that a conical interface is possible only when the
dielectric constant (or permeability) of the fluid  exceeds
a critical value $\epsilon_c=17.59$. For a fluid with $\epsilon>\epsilon_c$,
a  conical interface is possible at two  angles, one stable and one
unstable. We calculate the critical field required to sustain a drop with
stable conical tips.
Such a drop is  energetically favored  at sufficiently
 high field. Our
results
also apply to the formation of conical dimples when a pool of fluid
is placed  in a normal field. }

\vfil
\centerline{(Jan. 10, 1994)}
\eject
The study of the instability and breakup of a dielectric fluid drop
in a strong
electric field has a long history, due to its relevance to thunderstorms
and other practical problems such as ink-jet printing and
electrohydrodynamic atomization.
In his 1917 experiment, Zeleny\ref\rZeleny{J. Zeleny, {\sl Phys. Rev.}
 {\bf 10}, 1 (1917).}\
 showed that drops held at the end of capillary
tubes charged to high potential disintegrate due to the
formation of a  pointed end
from which a fine jet of fluid is ejected. Wilson and Taylor\ref\rWT{C. T.
R. Wilson and G. I. Taylor,
 {\sl Proc. Camb. Phil. Soc.}, {\bf 22}, 728 (1925).}\
 observed a similar phenomenon when
an uncharged soap bubble was subjected to a uniform electric field. Experiments
on free drops in strong fields have revealed comparable features\ref\rNolan{
J. J. Nolan, {\sl Proc. Roy. Irish Acad.}, {\bf 37}, 28 (1926).}\ref\rMacky{
W. A. Macky, {\sl Proc. Roy. Soc.} A, {\bf 133}, 565 (1931).}
\ref\rGarton{C. G. Garton and Z. Krasucki, {\sl Proc. Roy. Soc. Lond.} A, {\bf
280}, 211 (1964).}. At  low fields, the drop distorts to a slightly elongated
spheroid; as the field is increased
 above a threshold, the drop becomes unstable. As
it elongates, pointed ends develop from which
filaments of fluid are drawn
to the field plates.

Parallel experiments on ferrofluid drops
in magnetic fields were carried out much later.
In 1982, Bacri and Salin\ref\rBS{J. C. Bacri and D. Salin, {\sl J. Phys. Lett}
{\bf 43}, 649 (1982).}\ observed that above a critical field,
a ferrofluid drop became
unstable, jumping
from a prolate spheroidal shape
to a  much more elongated shape
with conical ends. The transition back to the less elongated spheroid
occured at a lower critical field,
 a  hysteresis characteristic of a first order phase transition.

 The theoretical understanding of this instability and shape transition
 is mostly
 based on the  approximation of spheroidal shape,
for which case analytical expressions for the
 field exist\rGarton\ref\rTaylor{G. I. Taylor, {\sl Proc. Roy Soc. Lond.} A,
{\bf 280}, 383 (1964).}\rBS. Taylor  analyzed the conditions  under which a
conical interface
can exist in equilibrium for a conducting fluid (with
dielectric constant $\epsilon=\infty$)\rTaylor, and found that a
conical interface can exist only at a particular half
angle of $49.3\degs$.
To analyze the full problem   without approximating the shape,
one needs to know the electric field inside a drop of arbitrary
shape, which requires numerical
calculations\ref\rBra{
P. R. Brazier-Smith, {\sl Phys. Fluids} {\bf 14}, 1 (1971).}\ref\rMik{
M. J. Miksis, {\sl Phys. Fluids} {\bf 24}, 1967 (1981).}\ref\rSher{
J. D. Sherwood, {\sl J. Fluid Mech.} {\bf 188}, 133 (1988).}\ref\rSer{
O. E. Sero-Guillaume, D. Zouaoui, D. Bernardin, and J. P. Brancher,
{\sl J. Fluid. Mech.} {\bf 241}, 215 (1992).}.

In this letter, we  take a semi-analytical approach to this problem.
We shall  discuss a dielectric drop
in an  electric field; the corresponding problem of a ferrofluid drop in
a magnetic field is completely analogous.
 We first generalize Taylor's argument for a conical interface to a
fluid with arbitrary dielectric constant $\epsilon$. We find a
critical value of the dielectric constant  $\epsilon_c=17.59$. For a
fluid with
$\epsilon < \epsilon_c$, no conical
interface can form. For a fluid with $\epsilon >\epsilon_c$,
a conical interface can exist in equilibrium at
two particular half angles, a larger angle $\thl$ and a smaller
angle $\ths$. In  the limiting case where
$\epsilon=\infty$, $\ths=0\degs$ and
 $\thl=49.3\degs$; the latter is  the angle found by Taylor.
We show that a conical interface at an angle $\thl$ is
unstable
while one at  $\ths$  can exist in stable equilibrium.
Thus Taylor's result does not correspond to any stable physical
situation, although such an angle can be observed
 as a transient\rZeleny\rTaylor. This result applies to a perfect conical
interface, but we are concerned instead with drops of finite
volume. We find that a stable drop with conical tips exists only
above a threshold field (which is a function of dielectric constant,
surface tension and volume). We calculate the energy of such a
drop and find that it is favored compared to the spheroidal shape
at a sufficiently high field.

Consider a fluid drop of volume $V\equiv 4\pi r_0^3/3$, placed
in a uniform applied field ${\bf E}_0$ oriented in the $z$ direction.
We assume that the dielectric constant of the fluid is $\epsilon$,
while that
of the surrounding medium may be taken as $1$.
 Due to the depolarization of the fluid,
 the total energy of the system is\ref\rLandau{L. D. Landau and E. M.
Lifshitz, {\sl Electrodynamics of Continuous Media}
 (Pergamon Press, 1960), chapter II.}
\eqn\eUt{U_t=\gamma S-\frac{1}{2}{\bf E}_0\cdot{\bf {\bf P}}\ ,}
where $\gamma$ is the surface tension, $S$ the surface area,
and ${\bf P}$ the dipole moment
of the drop.
 The  shape of the drop  is determined by the competition
between the surface tension, which favors a spherical shape,
 and depolarization,
which favors an elongated one. The equilibrium shape is  that which
minimizes $U_t$ with the
 total volume fixed.

Alternatively, the equilibrium shape can be determined by considering a
force
balance at the surface of the drop. The stress due to the
electric field is given
by the  stress tensor\rLandau
\eqn\eStr{\sigma_{ij}=\frac{\epsilon}{4\pi}E_iE_j-
         \frac{\epsilon}{8\pi}E^2\delta_{ij}\ ,}
which is discontinuous across the surface, giving rise to  a
normal pressure. Equilibrium is achieved when this
pressure is balanced by the surface tension plus the hydrodynamic pressure
difference. Therefore, the equilibrium condition is
\eqn\eFor{\frac{\epsilon-1}{8\pi}E^2+\frac{(\epsilon-1)^2}{8\pi}E_n^2
       +\rho_0=2\gamma {\cal H}\ ,}
where the first two terms  represent
normal pressure due to the field. Here $E$ is the field inside the drop,
$E_n$ its normal component,
 $\rho_0$ the hydrodynamic pressure difference,
and ${\cal H}$  the mean curvature.  The above equation relates the
local geometry of the drop to the local field. The problem is difficult
because
the local field is not known {\it a priori}, and must be determined from
the shape of the drop.

 We  first analyze the conditions  under which
a conical interface can exist by generalizing Taylor's argument for a
conducting fluid.
Consider a conical interface with a semi vertical angle $\tho$. We take
spherical coordinates ($r$,$\theta$,$\phi$), with the tip of the
cone at the origin and the region occupied by the fluid
 given by $0\leq\theta\leq\tho$ (see the insert to Fig.~1).
Simple geometry yields mean curvature ${\cal H}=\cot\tho/(2r)$, which diverges
as the tip is
approached. In equilibrium,  Eq.~\eFor\ then requires
that $E\propto 1/\sqrt{r}$.   The
potential inside and outside the cone satisfying Laplace's equation
 can be  written as
\eqn\eCone{\eqalign{&\Phi_{in}(r,\theta)=\sum_{\nu\geq 0}a_\nu r^\nu P_\nu
                                [\cos \theta], \tho\geq\theta\geq 0, \cr
            &\Phi_{out}(r,\theta)=\sum_{\nu\geq 0}b_\nu r^\nu P_\nu
                      [\cos (\pi-\theta)], \pi\geq\theta\geq\tho\ , \cr}}
where $P_\nu[x]$ is the Legendre function of the first kind.
Notice that for non-integer $\nu$, $P_\nu[x]$ is singular at $x=-1$, so
the argument $\pi-\theta$ is used in the outer region.
 The boundary conditions are
\eqn\eBound{\eqalign{&\Phi_1(r,\tho)=\Phi_2(r,\tho)\ ,\cr
              &\epsilon\frac{\partial \Phi_1}{\partial \theta}(r,\tho)=
                \frac{\partial \Phi_2}{\partial \theta}(r,\tho)\ .}}
Substituting Eq.~\eCone\ into the above equations leads to the following
eigenvalue
equation:
\eqn\eEig{P_{\nu}[\cos \tho]P_{\nu}'[-\cos\tho]+\epsilon P_{\nu}[
         -\cos\tho]P'_{\nu}[\cos\tho]=0\ .}
For a given $\epsilon$ and $\tho$, the above equation has a solution
$0<\nu < 1$, giving rise to a singular field $E\sim r^{\nu-1}$. Of course,
we are seeking solutions with $\nu=1/2$. There is a critical value
$\epc=17.59$ below which $\nu >1/2$ for all possible angles
 $0\leq\tho\leq \pi/2$, i.e.,
there is no solution with $\nu=1/2$. For $\epsilon >\epc$, there are two
angles  corresponding to $\nu=1/2$, $\thl$ and $\ths$.
 At the critical value $\epsilon=\epc$,
the minimum of $\nu$
reaches $1/2$ at $\tho=\thl=\ths=30\degs$.
 As $\epsilon\rightarrow\infty$,
$\ths\rightarrow 0\degs$ and $\thl\rightarrow\ 49.3\degs$ (see Fig.~2).

 The larger angle $\thl$  corresponds
to an unstable equilibrium, as can be seen by the following argument.   A
perturbation that slightly decreases the angle will lead to a decrease of
$\nu$, hence to a more singular field (see the insert to Fig.~2).
 This field  tends to
elongate the drop, thereby decreasing the angle $\tho$;
 thus the perturbation will
grow. Similarly, a perturbation that increases the angle will also grow.
 By the same  argument, we find that the conical interface corresponding
to $\ths$ is stable against small perturbations of $\tho$. Such an
interface can exist in stable equilibrium provided that the
external field is
sufficiently strong (see below).

We have seen that a drop with conical tips must have an electric field
that diverges as $1/\sqrt{r}$  with an
amplitude determined by  the force balance Eq.~(3). On the other hand,
this amplitude depends
on the applied field and the shape of the drop, and cannot be obtained by
the analysis of the infinite cone.
In order to find which external fields allow
conical tips, we now derive an approximate integral
equation for the field
that is very accurate for slender drops with conical tips of small
angle.  This integral
equation enables us to easily handle the singularity close to the tip,
and  numerically it is very efficient.

Consider a drop axi-symmetric  about the $z$ axis (the field direction).
 Choose  cylindrical coordinates $(r, \phi, z)$, and
parameterize the surface of the drop by $r=R(z)$, with $z\in [-l, l]$,
 where $l$ is the
semi long axis (see Fig.~1).
 Due to rotational symmetry, the electric field has only a radial and a
$z$ component,
$E_r(r,z)$ and $E_z(r,z)$, with $E_r(0,z)=0$.
 We expand the
field in a Taylor series in $r$, and keep only the leading term,
\eqn\eApp{\eqalign{&E_z(r,z)=E_1(z)\ ,\cr
                          &E_r(r,z)=rE_2(z)\ .\cr}}
Using this approximation, the equation $\nabla\cdot {\bf E}=0$ relates
$E_r$ to $E_z$,
\eqn\eRel{E_r(z)=-\frac{r}{2}\frac{ dE_1(z)}{dz}\ .}

Given the shape of the surface,
 the induced surface charge is related to $E_1(z)$ by
$\sigma (z)=P_n=\chi {\bf n}\cdot{\bf E}$, where
$\chi=(\epsilon-1)/4\pi$, and ${\bf n}$ is the unit normal to the drop surface.
 Simple geometry yields
\eqn\esurf{\sigma (z)=-\frac{\chi}{\sqrt{1+\dot R^2(z)}}\left[
     \dot R(z)E_1(z)+\frac{R(z)}{2}\dot E_1(z)\right]\ ,}
where $\dot f(z)\equiv df(z)/dz$.
To be self-consistent, we must have $E_0+E'(z)=E_1(z)$,
where $E'(z)$ is the field on the $z$ axis
due to the induced surface charge $\sigma (z)$.
Using Eq.~\esurf, this condition leads to the following
integral equation for the field
\eqn\eInt{E_1(z)+\pi\chi\int_{-l}^ldz' E_1(z') R^2(z')\frac{R^2(z')+
  3(z-z')R(z')\dot R(z')-2(z-z')^2}{[(z-z')^2+R^2(z')]^{5/2}}=E_0\ .}
The above equation allows us to approximately compute the field for a
drop of arbitrary shape. It is exact for a spheroid because
Eq.~\eApp\ is exact for a spheroid. For a drop with conical tips,
 $R(z)\approx (l-z)
\tan\ths$ as $z\rightarrow l$, and  Eq.~\eInt\ then has a singular solution
 of the form $E_1(z)\propto
(l-z)^{\nu-1}$.
 The cancellation of the singularity
on the left hand side of Eq.~\eInt\ as $z\rightarrow l$ leads to
the following equation relating $\tho$ and $\nu$,
\eqn\eexp{\pi\chi (1+\nu)A(\tho,\nu)\tan^2\tho =1\ ,}
with
\eqn\eamp{A(\tho, \nu)=\int_{-1}^{\infty}\frac{u(1+u)^{\nu}du}
        {[u^2+(1+u)^2\tan^2\tho]^{3/2}}\ .}
For typical values of $\epsilon>\epc$,
Eq.~\eexp\ determines $\ths$ with only
$\approx 0.1\%$ error compared to the exact result of Eq.~\eEig.
 Similarly, Eq.~\eApp\ determines the ratio of normal to
tangential component of the field with the same accuracy.

We now wish to use this approximation to find
the range of field values for which drops with stable conical tips
are expected to appear. We choose $R(z)$ to correspond to a
spheroid matched to two cones (see Fig.~1) of tip angle $\ths$.
 With the volume
fixed to be that of a sphere with radius $r_0$, only the aspect ratio
$\beta\equiv l/a$ can be varied,
where $a$ is the equatorial radius.

 For a  given aspect ratio, we solve
Eq.~\eInt\
numerically using the following procedure. We divide the drop
into a tip region $l-\Delta<|z|<l$ and a middle region $0<|z|<l-\Delta$.
For sufficiently small $\Delta$, the solution in the tip region must
have the form $E_1(z)=B(\beta,\epsilon)E_0\sqrt{r_0}/\sqrt{l-z}$,
 with the amplitude
$B(\beta,\epsilon)$ a
function of aspect ratio alone for fixed $\epsilon$.
 Equation~\eInt\ then couples $B$ with $E_1(z)$ in the
middle region. The resulting equation is solved  by discretizing
$z$ and solving the corresponding linear equations.
 Such a procedure enables us to compute the $E$ field, and
 hence $B(\beta,\epsilon)$, with high
accuracy.  In order that
the force balance hold in the tip region, we must have $E_1(z)=
\sqrt{\gamma} g(\epsilon)/\sqrt{l-z}$,
where $g(\epsilon)$ is a number that  depends only on the dielectric
constant $\epsilon$,
\eqn\egg{g(\epsilon)=(\cot\ths\cos\ths)^{1/2}\left[\frac{\epsilon-1}{8\pi}
    \left(\cos\ths+\frac{\tan\ths\sin\ths}{4}\right)^2+\frac

{\epsilon(\epsilon-1)}{8\pi}\left(\frac{3\sin\ths}{4}\right)^2\right]^{-1/2}\
.}
Thus the equation
\eqn\eampb{E_0B(\beta,\epsilon)=\sqrt{\frac{\gamma}{r_0}}g(\epsilon)}
fixes the aspect ratio as a function of the  applied field.

Note that
the right-hand side of Eq.~\eampb\ is independent of the aspect ratio.
For a given
$\epsilon$, the function $B(\beta,\epsilon)$ has a maximum at
$\beta_{max}$. Thus only for
$E_0>E_{min}\equiv\sqrt{\gamma/r_0}g(\epsilon)
/B(\beta_{max},\epsilon)$
is there a solution to Eq.~\eampb, i.e., only for these fields can drops
with conical tips exist.

We recall that Bacri and Salin discussed two solutions--a slightly elongated
and a much more elongated spheroid. At an upper critical field $E^{BS}_>$,
the less elongated solution becomes unstable to the more elongated one,
this latter in turn loses stability to the less elongated solution at
a lower critical field $E^{BS}_< $. As $\epsilon\rightarrow\infty$,
$E^{BS}_<\rightarrow 0$ while $E^{BS}_>$ approaches a non-zero constant.
We find that $E_{min}<E^{BS}_<$ for all $\epsilon$ (see Fig.~3), which
suggests that the more elongated shape discussed by Bacri and Salin will
not be spheroidal but will instead have singular tips.

We can address this problem by comparing the energies of the two solutions.
The numerical solution of Eq.~\eInt\ can  be used to compute the
 total energy as a function of aspect ratio for a drop with conical tips
for a given $E_0$. The equilibrium shape is determined by the minimum of
the energy curve. From Eq.~\eUt,
\eqn\etotal{U_t=4\pi\gamma\int_0^l R(z)\sqrt{1+\dot R^2(z)}
    dz-\pi\chi E_0\int_0^lR^2(z)E_1(z)dz\ .}
We find that  for $E_0>E^{BS}_<$, a
drop with conical tips  has  a lower energy and a larger aspect ratio
compared to that of the more elongated
spheroid. Therefore,
it is energetically favorable to have a drop with conical tips
at high field.
 However,
 the difference in the total energy is quite small, typically
$\sim 1\%$. We believe that this is the reason why
Bacri and Salin obtained a quite accurate phase diagram using
purely spheroidal drops.

We have also done calculations using  a different one parameter class of
curves, with a fixed tip angle $\ths$, and
 $R(z)=c_1+c_2z^2+c_3|z|^3$.
The results are entirely analogous.

Our analysis of the formation of conical interfaces is consistent with
 experimental observations. The unstable angle $\thl$ was observed in
Zeleny's experiment in which  water and glycerine broke  from
capillary tubes\rZeleny. The measured angles agree quite well with our
calculated values of
$\thl=47.2\degs $ for water and $\thl=45.1\degs $ for glycerine.
The instability of $\thl$ was clearly demonstrated by Taylor's
experiment, in which strong oscillations of an oil-water interface
were observed as $\thl$ was
approached\rTaylor.

 For the stable angle $\ths$,
our prediction is consistent with Bacri and Salin's experiment
(using their estimate for the permeability).
Such an angle was
also observed  when a pool of ferrofluid
became unstable and developed dimples and spikes in a normal
 field\ref\rRosen{M. D. Cowley and
R. E. Rosensweig, {\sl J. Fluid Mech.} {\bf 30}, 671 (1967)}. Our
analysis provides a way of measuring the permeability
of the fluid by measuring the angle  of the spikes\ref\rnote{The conical
 interfaces observed in these experiments
are not infinitely sharp, of course,  due to the  cutoff by the saturation
effect.}. For the dielectric drops, the association of the shape
transition with the formation of conical tips clearly indicates
why the elongated shape is unstable to breakup as $\epsilon\rightarrow
\infty$; this point was left obscure by the earlier spheroidal calculations.
\vfil
\eject
\centerline{{\bf Acknowledgement}}
\medskip
H.L. thanks Tom Witten for many helpful discussions. This work was
supported by the  National Science Foundation
 through grant No. DMR-9208527 and through the Materials Research
Laboratory at the University of Chicago.
 T.C.H is grateful to the NSF for the support
of this research through the PYI program, grant DMR-9057156.
Acknowledgement is made to the Donors of the Petroleum Research
Fund, administered by the American Chemical Society, for the partial
support of this research.

\vfill\eject\immediate\closeout\rfile
\centerline{{\bf References}}\bigskip{
\catcode`\@=11\escapechar=` %
\input refs.tmp\vfill\eject}
\caption
\fig{1}{A drop with conical ends constructed by matching
a spheroid with two cones of angle $\ths$.
  The shape of the drop is parameterized by the curve $r=R(z)$, where the
$z$ axis is parallel to the applied field. Insert: a perfect conical
interface with semi-vertical angle $\theta_0$.}
\fig{2}{Angles $\ths$  and $\thl$
 as functions of dielectric constant
$\epsilon$ for $\epsilon>\epsilon_c$; the dashed vertical line
corresponds to $\epsilon=\epc$. Insert: Exponent $\nu$
as a function of the semi-vertical angle $\tho$
for $\epsilon>\epc$.  The dashed horizontal line corresponds to
$\nu=1/2$.}
\fig{3}{The minimum field $E_{min}$ (in unit of $\sqrt{\gamma/r_0}$)
required to sustain a drop with conical tips, as a function of
the dielectric constant $\epsilon$. The dotted and dashed lines
correspond to the lower and upper critical fields for the shape
transition calculated using
Bacri and Salin's method, which assumes purely spheroidal shapes.}

\end